\documentclass[twocolumn,aps,prd,epsf]{revtex4}
\usepackage{epsfig}
\usepackage{dcolumn}

\newcommand{\be}{\begin{equation}}
\newcommand{\ee}{\end{equation}}
\newcommand{\bea}{\begin{eqnarray}}
\newcommand{\eea}{\end{eqnarray}}

\begin{document}

\title{ The large degeneracy of excited hadrons and quark models }
\author{P. Bicudo}
\affiliation{Dep. F\'{\i}sica and CFTP, Instituto Superior T\'ecnico,
Av. Rovisco Pais, 1049-001 Lisboa, Portugal}
\begin{abstract}
The pattern of a large approximate degeneracy of the excited hadron spectra
(larger than the chiral restoration degeneracy) is present in the recent 
experimental report of Bugg. 
Here we try to model this degeneracy with state of the art quark models.
We review how the Coulomb Gauge chiral invariant and confining Bethe-Salpeter equation simplifies in 
the case of very excited quark-antiquark mesons, including angular or radial excitations, to a 
Salpeter equation with an ultrarelativistic kinetic energy with the spin-independent part of the 
potential. The resulting meson spectrum is solved, and the excited chiral restoration is recovered,
for all mesons with $J > 0$.
Applying the ultrarelativistic simplification to a linear equal-time potential, linear Regge 
trajectories are obtained, for both angular and radial excitations. The spectrum is also compared 
with the semi-classical Bohr-Sommerfeld quantization relation.
However the excited angular and radial spectra do not coincide exactly. We then search, with the
classical Bertrand theorem, for central potentials producing always classical closed orbits
with the ultrarelativistic kinetic energy. We find that no such potential exists, and this
implies that no exact larger degeneracy can be obtained in our equal-time framework, with a single principal 
quantum number comparable to the non-relativistic Coulomb or harmonic oscillator potentials. 
Nevertheless we find plausible that the large experimental approximate degeneracy will be modelled 
in the future by quark models beyond the present state of the art.
\end{abstract}

\maketitle

\twocolumngrid

%
%
%
%
%
%
%
%
%
%
\section{Introduction}

In the recent report of Bugg 
\cite{Bugg:2004xu,Afonin:2006vi,Glozman:2007ek,fig1afon}
a large degeneracy emerges from the spectra of the angularly and radially excited resonances 
produced in $p \bar p$ annihilation by the Crystal Ball collaboration at LEAR 
in CERN
\cite{Aker:1992ny}.
This degeneracy may be the third remarkable pattern of the excited 
spectra of hadronic resonances. 

A long time ago, Chew an Fautschi remarked the existence
of linear Regge trajectories 
\cite{Chew}
for angularly excited mesons,
and baryons
\cite{Fiore:2004xb,Yao:2006px}.
A similar linear aligning of excited resonances was also reported for radial excitations
\cite{Anisovich:2001pn}.

Recently, Glozman {\em et al.} 
\cite{Glozman:2007ek,Wagenbrunn:2007ie,Glozman}
have been systematically researching the degeneracy 
of chiral partners in excited resonances, both in models
and in lattice QCD.
   Le Yaouanc {\em et al.} 
\cite{Yaouanc}
and PB {\em et al.} 
\cite{Bicudo_thesis,Bicudo_hvlt}
developed a model of spontaneous
   symmetry breaking. Viewed retrospectively, this already 
   included the degeneracy of chiral partners in the limit of 
   high radial or angular excitations
\cite{Malheiro},
both in light-light and in heavy-light hadrons. 
This earlier work was 
   based on the Bogoliubov transformation and used the Coulomb
   gauge or the local coordinate gauge QCD truncations. Here the
   formalism will be extended to include spin-dependent 
   interactions explicitly.

The recent data on highly excited mesons, observed by LEAR at CERN
\cite{Bugg:2004xu,fig1afon,Aker:1992ny}
and of excited baryons observed by the Crystal Barrel collaboration at ELSA 
\cite{Thoma},  
stress the interest of possible patterns in the excited hadronic resonance 
spectra. While we still have to wait for new experiments focused on excited
mesonic resonances to confirm the report of Bugg, say in PANDA, GLUEX, or BESIII,
it is important to research theoretical models of the excited hadrons.

Here we address the question, is it possible to build an equal-time
quark model, with linear trajectories, with
excited chiral symmetry, and, also, with a principal quantum number?
We adopt the framework of the Coulomb gauge confinement, of the mass
gap equation, and of the equal-time Bethe-Salpeter equation.
In Section II we review and expand earlier work on 
chiral symmetry breaking and mesonic boundstates, 
and show in detail the equations for the simplest potential.
We also show how for excited states (and for $J > 0$) the Bethe-Salpeter
equation simplifies to a Schr\"odinger-like Salpeter equation, with
ultrarelativistic (massless) kinetic energies and a chiral symmetric 
equal-time potential.
In Section III we solve the equation with the method of the double diagonalization 
of the equal time hamiltonian and show how linear equal-time potentials
and massless quarks produce linear Regge trajectories, both for angular and radial 
excitations.
We also compare them with the Bohr-Sommerfeld semi-classical quantization. 
We then address in Section IV the large degeneracy, where both radial and angular
excitations are degenerate. Extending to ultrarelativistic particles the techniques
of the classical Bertrand theorem on closed orbits, we verify that 
no instantaneous 2-body potential may exactly produce the desired
large degeneracy. Nevertleless we present plausible solutions to this modelling
problem in the conclusion, Section V.

\section{Quark mass gap and boundstates in equal time}

We first review earlier work on chiral symmetry breaking with 
equal-time confining quark-quark potentials, and show the example of
the simplest possible model of this class of potentials, 
which continues to be explored
\cite{Bicudo:2005de}. 
Importantly, the hamiltonian of this model can be approximately derived from QCD, 
\begin{eqnarray}
&&H=\int\, d^3x \left[ \psi^{\dag}( x) \;(m_0\beta -i{\vec{\alpha}
\cdot \vec{\nabla}} )\;\psi( x)\;+
{ 1\over 2} g^2 \int d^4y\, \
\right.
\nonumber \\
&&
\overline{\psi}( x)
\gamma^\mu{\lambda^a \over 2}\psi ( x)  
\langle A_\mu^a(x) A_\nu^b(y) \rangle
\;\overline{\psi}( y)
\gamma^\nu{\lambda^b \over 2}
 \psi( y)  \ + \ \cdots
\label{hamilt}
\end{eqnarray}
up to the first cumulant order, of two gluons
\cite{Bicudo_hvlt,Dosch,Kalashnikova,Nefediev}.
In the modified coordinate gauge the cumulant is,
\begin{equation}
g^2 \langle A_\mu^a(x) A_\nu^b(y) \rangle
\simeq-{3 \over 4} \delta_{ab} g_{\mu 0} 
g_{\nu 0}
\left[K_0^3({\bf x}-{\bf y})^2-U\right]
\label{potential}
\end{equation}
and this is a simple density-density harmonic effective 
confining interaction. $m_0$ is the current mass of the quark.
The infrared constant $U$ confines the quarks but the 
meson spectrum is completely insensitive to it. 
The important parameter is the potential strength $K_0$, 
the only physical scale in the interaction, and all results
can be expressed in units of $K_0$. A reasonable
fit of the hadron spectra is achieved with
$K_0 \simeq 0.3\pm 0.05$ GeV.

%
\begin{figure}[t]
\includegraphics[width=0.80\columnwidth]{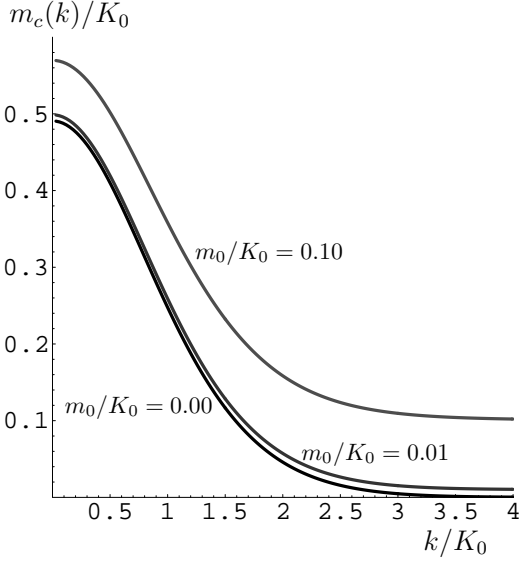}
\caption{
The constituent quark masses $m_c(k)$, solutions of the mass gap equation,
for different current quark masses $m_0$. 
}\label{mass solution}
\end{figure}

The relativistic invariant Dirac-Feynman propagators
\cite{Yaouanc}, 
can be decomposed in the quark and antiquark Bethe-Goldstone 
propagators
\cite{Bicudo_scapuz},
used in the formalism of non-relativistic quark models,
\FL
\begin{eqnarray}
{\cal S}_{Dirac}(k_0,\vec{k})
&=& {i \over \not k -m +i \epsilon}
\nonumber \\
&=& {i \over k_0 -E(k) +i \epsilon} \
\sum_su_su^{\dagger}_s \beta
\nonumber \\
&& - {i \over -k_0 -E(k) +i \epsilon} \
\sum_sv_sv^{\dagger}_s \beta \ ,
\nonumber \\
u_s({\bf k})&=& \left[
\sqrt{ 1+S \over 2} + \sqrt{1-S \over 2} \widehat k \cdot \vec \sigma \gamma_5
\right]u_s(0)  \ ,
\nonumber \\
v_s({\bf k})&=& \left[
\sqrt{ 1+S \over 2} - \sqrt{1-S \over 2} \widehat k \cdot \vec \sigma \gamma_5
\right]v_s(0)  \ ,
\nonumber \\
&=& -i \sigma_2 \gamma_5 u_s^*({\bf k}) \ ,
\label{propagators}
\end{eqnarray}
where $S=\sin(\varphi)={m_c\over \sqrt{k^2+m_c^2}} \ , 
\ C=\cos(\varphi)={k\over \sqrt{k^2+m_c^2}}$ and $\varphi$ is a chiral angle.
In the non condensed vacuum, $\varphi$ is equal to $\arctan{m_0 \over k}$.
In the physical vacuum, the constituent quark mass $m_c(k)$, or the
chiral angle $\varphi(k)=\arctan{m_c(k) \over k}$, is a variational function
which is determined by the mass gap equation. We anticipate examples of 
solutions, for different light current quark masses $m_0$, 
depicted in Fig. \ref{mass solution}.

There are three equivalent methods to derive the mass gap equation
for the true and stable vacuum, where constituent quarks acquire 
the constituent mass
\cite{Bicudo:2005de}. 
One method consists in assuming a quark-antiquark $^3P_0$ 
condensed vacuum, and in minimizing the vacuum energy density. 
A second method consists in rotating the quark and antiquark 
fields with a Bogoliubov-Valatin canonical transformation 
to diagonalize the terms in the hamiltonian with two   
quark or antiquark second quantized fields. 
A third method consists in solving the Schwinger-Dyson 
equations for the propagators. Any of these methods
lead to the same mass gap equation and quark 
dispersion relation. 
Here we replace the propagator
of eq. (\ref{propagators}) in the Schwinger-Dyson equation, 
\begin{eqnarray}
\label{2 eqs}
&&0 = u_s^\dagger(k) \left\{k \widehat k \cdot \vec \alpha + m_0 \beta
-\int {d w' \over 2 \pi} {d^3k' \over (2\pi)^3}
i V(k-k') \right.
\nonumber \\
&&\left. \sum_{s'} \left[ { u(k')_{s'}u^{\dagger}(k')_{s'} 
 \over w'-E(k') +i\epsilon}
-{ v(k')_{s'}v^{\dagger}(k')_{s'} 
  \over -w'-E(k')+i\epsilon} \right]
\right\} v_{s''}(k) \  \
\nonumber \\
&&E(k) = u_s^\dagger(k) \left\{k \widehat k \cdot \vec \alpha + m_0 \beta
-\int {d w' \over 2 \pi} {d^3k' \over (2\pi)^3}
i V(k-k')  \right.
\nonumber \\
&&\left. \sum_{s'} \left[ { u(k')_{s'}u^{\dagger}(k')_{s'} 
 \over w'-E(k') +i\epsilon}
-{   v(k')_{s'}v^{\dagger}(k')_{s'}  
 \over -w'-E(k')+i\epsilon} \right]
\right\} u_s(k),
\end{eqnarray}
where, with the simple density-density harmonic interaction
\cite{Yaouanc}, the integral of the potential is a laplacian 
and the mass gap equation and the quark energy are finally,
\begin{eqnarray}
\label{mass gap}
\Delta \varphi(k)  &=& 2 k S(k) -2 m_0 C(k) - { 2 S(k) C(k) \over k^2 }  
\\ \nonumber 
E(k)&=& k C(k) + m_0 S(k) - { {\varphi'(k) }^2 \over 2 } - { C(k)^2 \over k^2 } 
+{ U \over 2} \ .
\end{eqnarray}
Numerically, this equation is a non-linear ordinary differential
equation. It can be solved with the Runge-Kutta and shooting method.
Examples of solutions for the current quark mass $m_c(k)= k \tan \varphi$, 
for different current quark masses $m_0$, 
are depicted in Fig. \ref{mass solution}.

%
%
\begin{table}[t]
\begin{ruledtabular}
\begin{tabular}{c|c}
& $V^{++}=V^{--}$  \\ \hline
spin-indep. & $- {d^2 \over dk^2 } + { {\bf L}^2 \over k^2 } + 
{1 \over 4} \left( {\varphi'_q}^2 + {\varphi'_{\bar q}}^2 \right) 
+ {  1 \over k^2} \left( {\cal G}_q +{\cal G}_{\bar q}  \right) -U $  \\ 
spin-spin & $ {4 \over 3 k^2} {\cal G}_q {\cal G}_{\bar q} {\bf S}_q \cdot {\bf S}_{\bar q} $  \\ 
spin-orbit & $ {1 \over  k^2} \left[ \left( {\cal G}_q + 
{\cal G}_{\bar q} \right) \left( {\bf S}_q +{\bf S}_{\bar q}\right) 
+\left( {\cal G}_q - {\cal G}_{\bar q} \right) \left( {\bf S}_q -{\bf S}_{\bar q}\right)  \right]
\cdot {\bf L} $  \\ 
tensor & $ -{2 \over  k^2} {\cal G}_q {\cal G}_{\bar q} 
\left[ ({\bf S}_q \cdot \hat k ) ({\bf S}_{\bar q} \cdot \hat k )
-{1 \over 3} {\bf S}_q \cdot {\bf S}_{\bar q} \right] $ \\ \hline
& $V^{+-}=V^{-+}$ \\ \hline
spin-indep. & $0$  \\ 
spin-spin & $ -{4 \over 3} \left[ {1\over 2} {\varphi'_q} {\varphi'_{\bar q}} + 
{1\over k^2} {\cal C}_q {\cal C}_{\bar q}  \right]
{\bf S}_q \cdot {\bf S}_{\bar q} $  \\ 
spin-orbit & $0$  \\ 
tensor & $ \left[ -2 {\varphi'_q} {\varphi'_{\bar q}} + 
{2\over k^2} {\cal C}_q {\cal C}_{\bar q}  \right]
\left[ ({\bf S}_q \cdot \hat k ) ({\bf S}_{\bar q} \cdot \hat k )
-{1 \over 3} {\bf S}_q \cdot {\bf S}_{\bar q} \right] $
\end{tabular}
\end{ruledtabular}
\caption{\label{spin dependent} 
The positive and negative energy spin-independent, spin-spin, spin-orbit and tensor 
potentials, computed exactly in the framework of the simple density-density harmonic model
of eq. (\ref{potential}).
$\varphi'(k)$, ${\cal C}(k)$ and ${\cal G}(k)= 1 - S(k) $ are all functions of the constituent 
quark(antiquark) mass.
}
\end{table}

The Salpeter-RPA equations for a meson (a colour singlet
quark-antiquark bound state) can be derived from the Lippman-Schwinger
equations for a quark and an antiquark, or replacing the propagator
of eq. (\ref{propagators}) in the Bethe-Salpeter equation. In either way, one gets
\cite{Bicudo_scapuz}
\FL
\begin{eqnarray}
\label{homo sal}
\phi^+(k,P) &=& { u^\dagger(k_1) \chi(k,P)  v(k_2) 
\over +M(P)-E(k_1)-E(k_2) }
\nonumber \\
{\phi^-}^t(k,P) &=& { v^\dagger(k_1) \chi(k,P) u(k_2)
\over -M(P)-E(k_1)-E(k_2)}
\nonumber \\
\chi(k,P) &=&
\int {d^3k' \over (2\pi)^3} V(k-k') \left[ 
u(k'_1)\phi^+(k',P)v^\dagger(k'_2) \right.
\nonumber \\
&&\left. +v(k'_1){\phi^-}^t(k',P) u^\dagger(k'_2)\right] 
\end{eqnarray}
where $k_1=k+{P \over 2} \ , \ k_2=k-{P \over 2}$ and $P$ is
the total momentum of the meson.

The Salpeter-RPA equations of PB et al. 
\cite{Bicudo_thesis}
and of Llanes-Estrada et al. 
\cite{Llanes-Estrada_thesis}
are obtained deriving the equation for the positive energy wavefunction
$\phi^+$ and for the negative energy wavefunction $\phi^-$. 
Solving for $\chi$, one gets the Salpeter equations of Le Yaouanc et al.
\cite{Yaouanc}.
This results in four potentials 
$V^{\alpha \beta}$ respectively coupling $\nu^\alpha=r \phi^\alpha$ to 
$\nu^\beta$, in the boundstate Salpeter equation,
\begin{equation}
\left\{
\begin{array}{rrr}
\left(2 T + V^{++} \right) \, \nu^+ \  + & V^{+-} \ \ \nu^- =& M \nu^+ \\
V^{-+} \ \  \nu^+ \ + & \left( 2 T + V^{--} \right) \, \nu^- =& - M \nu^- 
\end{array}
\right. .
\label{decoupled}
\end{equation}
The relativistic equal time equations have the double of coupled
equations than the Schr\"odinger equation. The negative energy component 
$\nu^-$ is smaller than the positive energy component by a factor of the
order of $1/M$ in units ok $K_0=1$. Thus when $M$ is large, and this is the
case for most excited mesons, the negative energy components can be neglected
and the Salpeter equation simplifies to a Schr\"odinger equation.

Importantly, the potentials $V^{++}=V^{--}$ and $V^{+-}=V^{-+}$ include 
the usual spin-tensor potentials
\cite{Bicudo_baryon}, 
produced by the Pauli $\vec \sigma$ matrices in the spinors of eq. (\ref{propagators}).
They are detailed explicitly in Table \ref{spin dependent}. 
Because we are interested in highly excited states, where both $\langle r \rangle $ and 
$\langle k \rangle $ are large, we consider the limit where ${m_c \over k} \rightarrow 0$. 
This implies that the potentials, used in Table \ref{spin dependent}, $\varphi'(k)\rightarrow 0$, 
${\cal C}(k)\rightarrow 1$ and 
${\cal G}(k)= 1 - S(k)  \rightarrow 1 $. Then using the textbook matrix elements of the 
spin-tensor potentials, the boundstate Salpeter equation decouples in two different equations 
depending only on $\mathbf J$ and not explicitly on $\mathbf L$ or $\mathbf S$.
Without the chiral degeneracy there would be four different resonances for each $j$,
one with $s=0$ and $j=l$ and three with $s=1$ and $j=l-1, \, l, \, l+1$. With the
chiral degeneracy we get only two different equations, one for 
$j \geq 0$ with,
\begin{equation}
\label{J g 0}
\left\{ 
\begin{array}{cl}
2 T + V^{++}=& -{d^2 \over d k ^2} +2 k  - {1\over k^2} + {j(j+1) \over k^2} \\ 
 V^{+-}=& {1\over k^2} \\
\end{array} \right. \ ,
\end{equation}
and another for $j \geq 1$ with,
\begin{equation}
\label{J g 1}
\left\{ 
\begin{array}{cl}
2 T + V^{++}=& -{d^2 \over d k ^2} +2 k  - {2\over k^2} + {j(j+1) \over k^2}\\ 
 V^{+-}=& {0\over k^2} \\
\end{array} \right. \ .
\end{equation}
Thus states with different $l$ and equal $j$, i.e. with different parity are
degenerate, and chiral symmetry is restored.
This chiral degeneracy applies to all angular momenta, except for $j=0$.
In Table 
\ref{tachyon+meson} we show the masses of the different light-light mesonic
solutions of eqs.  (\ref{J g 0}) and (\ref{J g 1}), including tachyons, 
corresponding to the limit where $m_c<<k$.

%
%
\begin{table}[t]
\begin{ruledtabular}
\begin{tabular}{c|ccrrrrrrrr}
n & Pse & Sca & j=1 & j=1 & j=2	& j=2 & j=3 & j=3    \\ \hline
0            &  $2 \times 10^{-1} i \over m^2$ &  $3 \times 10^{-2}  i \over m^2$ & 3.71 & 4.59 & 6.15 & 6.45 & 7.65 &  7.84   \\ 
1            &  $2 \times 10^{-3} i \over m^2$ &  $3 \times 10^{-4}  i \over m^2$ & 6.49 & 7.15 & 8.43 & 8.69 & 9.72 & 9.89   \\ 
2            &  $2 \times 10^{-5} i \over m^2$ &  $3 \times 10^{-6} i \over m^2$ & 8.76 & 9.32 & 10.45 & 10.68 & 11.61 & 11.76  \\ 
3            &  $2 \times 10^{-7} i \over m^2$ &  $3 \times 10^{-8} i \over m^2$ & 10.77 & 11.27 & 12.30 & 12.51 & 13.38 & 13.52   \\  
4            &  $2 \times 10^{-9} i \over m^2$ &  $3 \times 10^{-10} i \over m^2$ & 12.61 & 13.08 & 14.05 & 14.25 & 15.12 & 15.26   \\  
\end{tabular}
\end{ruledtabular}
\caption{\label{tachyon+meson} 
Masses of the first angular and radial excitations of the different 
light-light tachyons and mesons in the chiral limit of $m<<k$. 
Each column includes both positive and 
negative parity degenerate states, except for the pseudoscalar 
and scalar tachyonic states. The meson masses are separated in two different 
families with the same $J$ because two different Salpeter equations 
(\ref{J g 0}) and (\ref{J g 1}) exist for each $J$.
}
\end{table}

The $j=0$ case is a subtle one, because neglecting the mass
$m_c$ is equivalent to consider the chiral limit, and this
is equivalent to change from the physical vacuum to the
chiral invariant vacuum which is a false, unstable vacuum. 
A detailed inspection shows that the different potentials 
$ -{d^2 \over d k ^2} \, , \
2 k \, , \ {1\over k^2} $ are bound from below and 
positive definite in the sense that all their eigenvalues
are positive. However $-{1\over k^2} $ is unbound from
below. It turns out that for $j=0$ all the solutions of  
eq. (\ref{J g 0}), including all radial excitations,
are tachyons
\cite{Bicudo:2006dn},
relevant for the structure of the chiral invariant false vacuum of QCD
\cite{Hiller},
corresponding to a different solution of the mass gap equation $m_c=0$.
Even when a very small regularizing quark mass $m_c$ is assumed, constant
for simplicity, 
the tachyons persist. 
This is confirmed by the numerical solutions of the regularized Salpeter equation,
shown in Table \ref{tachyon+meson} . 
Technically, in the $j=0$ case, it is necessary to rescale
the momentum and mass,
\begin{eqnarray}
k / m_c &\rightarrow& {k'} \ ,
\nonumber \\
M \, {m_c}^2 &\rightarrow& M' \ ,
\end{eqnarray}
where any finite solution $M'$ in fact 
corresponds to a large mass $M=M' /m_c^2 $, and
where a wave-function with a finite $k'$ corresponds 
to a wave-function with small momentum $k=k'\, m_c$.
A long time ago Le Yaouanc et al. 
\cite{Yaouanc}
showed that in the chiral limit the pseudoscalar and the scalar possess tachyonic solutions. 
Very recently PB showed that this number of tachyons is infinite
\cite{Bicudo:2006dn}.
Only with a finite $m_c$ quark mass, do the scalar and pseudoscalar mesons have positive masses. 
But then the excited scalar and pseudoscalar states are not degenerate, in contradistinction 
with the chiral degeneracy of the excited mesons with $j>0$. 

For excited mesons with $j>0$, the spectrum in table 
\ref{tachyon+meson} is very well aproximated by the solutions of the 
pair of Schr\"odinger-like equations,
\begin{eqnarray}
\left[-{d^2 \over d k ^2} +2 k  - {1\over k^2} + {j(j+1) \over k^2} \right] \nu(k) &=& M \nu(k) 
\label{-1/k2}
\\ 
\left[-{d^2 \over d k ^2} +2 k  - {2\over k^2} + {j(j+1) \over k^2}\right] \nu(k) &=& M \nu(k)
\label{-2/k2}
\end{eqnarray}
obtained when the negative energy component $\nu^-$ are neglected in eqs (\ref{J g 0}) and (\ref{J g 1}).

\section{Linear Regge trajectories and semi-classical quantization}

The simple quadratic model is now extended to a linear potential
\cite{Llanes-Estrada_thesis,Bicudo_KN},
to get the linear Regge trajectories. It is well known, both from
the quark modelling of the hadronic spectra, and from Lattice QCD,
that the long range confining quark potential is linear. Notice that 
Szczepaniak {\em et. al.}
\cite{Szczepaniak}, 
in the coulomb gauge, were able to derive 
from QCD the linear potential.
Continuing with the limit where ${m \over k} \rightarrow 0$, we assume
the radial equation,
\bea
p &=&  \sqrt{ - {d^2 \over d r^2} + { j(j+1) \over r^2 } }
\nonumber \\
\left( p  + \sigma \ r  \right) \nu(r) &=& E \nu(r) \ ,
\label{ultrarelativistic}
\eea
where the negative energy components (vanishing with $1 \over M$) were neglected,
and where the terms similar to the terms $-1 \over k^2$ or $-2 \over k^2$
of eq. (\ref{-1/k2}) or  eq. (\ref{-2/k2})were also neglected. We only expect these  
approximations to be reasonable for highly excited states, and for $j > 0$ 
because the number of excited false vacua remains infinite 
\cite{Bicudo_rep}.

Then we have to solve a Salpeter equation (or Schr\"odinger equation with
ultrarelativistic kinetic energy) except that the spherical angular
momentum $l$ is now replaced by the total angular momentum $j$,
in the centrifugal barier. 
Eq.(\ref{ultrarelativistic}) is solved with the method of the double 
diagonalization of the equal time hamiltonian. Using finite differences, 
first we diagonalize the bounded from below operator 
$- {d^2 \over d r^2} + { j(j+1) \over r^2 }$ . Then we apply the square root.
After returning to the original position space basis, we diagonalize the full hamiltonian.
This provides us automatically with the full spectrum including the angularly and
radially excited states.

%
%
\begin{table}[t]
\begin{ruledtabular}
\begin{tabular}{r|rrrrrrrrr}
$j$& $n$=0& $n$=1& $n$=2& $n$=3& $n$=4& $n$=5& $n$=6& $n$=7& $n$=8 \\ \hline
 0& 3.16& 4.71& 5.89& 6.87& 7.73& 8.51& 9.21& 9.87& 10.49\\
 1& 4.22& 5.46& 6.48& 7.38& 8.17& 8.90& 9.58& 10.21& 10.81\\
 2& 5.08& 6.13& 7.05& 7.87& 8.61& 9.30& 9.95& 10.56& 11.13\\
 3& 5.81& 6.74& 7.58& 8.34& 9.04& 9.70& 10.31& 10.90& 11.45\\
 4& 6.46& 7.31& 8.08& 8.79& 9.45& 10.08& 10.67& 11.23& 11.77\\
 5& 7.05& 7.83& 8.55& 9.22& 9.86& 10.45& 11.02& 11.56& 12.08\\
 6& 7.60& 8.32& 9.00& 9.64& 10.24& 10.82& 11.36& 11.89& 12.39\\
 7& 8.11& 8.79& 9.43& 10.04& 10.62& 11.17& 11.70& 12.21& 12.70\\
 8& 8.59& 9.23& 9.84& 10.43& 10.98& 11.51& 12.03& 12.52& 12.99\\
 9& 9.04& 9.65& 10.24& 10.80& 11.33& 11.85& 12.34& 12.82& 13.29\\
10& 9.47& 10.06& 10.62& 11.16& 11.68& 12.18& 12.66& 13.12& 13.58\\
11& 9.88& 10.45& 10.99& 11.51& 12.01& 12.49& 12.96& 13.42& 13.86\\
12& 10.28& 10.82& 11.35& 11.85& 12.34& 12.81& 13.26& 13.71& 14.14\\
13& 10.66& 11.19& 11.69& 12.18& 12.65& 13.11& 13.56& 13.99& 14.41\\ 
\end{tabular}
\end{ruledtabular}
\caption{\label{ultrarelativistic meson} 
Masses of the light-light mesons, in dimensionless units of $\sigma=1$, 
computed with the ultrarelativistic equal time chiral degenerate
Schr\"odinger equation (\ref{ultrarelativistic}). 
The $j=0$ mesons are distant from the experimental spectrum, but
chiral degeneracy is theoretically plausible for the very excited mesons.
}
\end{table}

\begin{figure}[t]
\includegraphics[width=1.05\columnwidth]{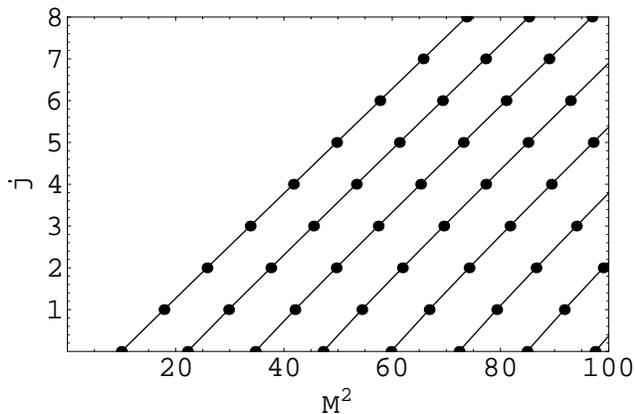}
\caption{
We show the quasi-linear Regge trajectories, of $j$ as a function of $M^2$.
Each line corresponds to a fixed radial $n$, increasing from left to right.
The $M$ are the masses of the light-light mesons, in dimensionless units of $\sigma=1$, 
computed with the ultrarelativistic equal time  chiral degenerate
Schr\"odinger equation (\ref{ultrarelativistic}). 
}\label{angular Regges}
\end{figure}

\begin{figure}[t]
\includegraphics[width=1.05\columnwidth]{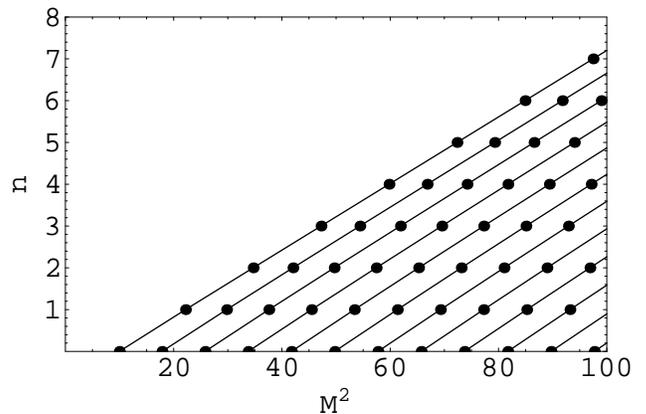}
\caption{
We show we show the quasi-linear Regge trajectories, of $n$ as a function of $M^2$.
Each line corresponds to a fixed angular $j$, increasing from left to right.
The $M$ are the masses of the light-light mesons, in dimensionless units of $\sigma=1$, 
computed with the ultrarelativistic equal time  chiral degenerate
Schr\"odinger equation (\ref{ultrarelativistic}). 
}\label{radial Regges}
\end{figure}

The results are shown in Table \ref{ultrarelativistic meson}. In
Fig. \ref{angular Regges} and in Fig. \ref{radial Regges} we 
graphically demonstrate that the angular excitations and the
radial excitations of this simple spectrum
are disposed in linear Regge trajectories,
\bea
j &\simeq&  \alpha_0 + \alpha M^2 \ ,
\nonumber \\
j &\simeq&  \beta_0 + \beta  M^2 \ .
\label{Regge slopes}
\eea
This agrees qualitatively with the experimental spectrum
\cite{Bugg:2004xu}, where the linear Regge trajectories are 
also present.

Interestingly, with the semi-classical Bohr-Sommerfeld quantization relation
\be
\oint p \ dq \simeq n \, h  \ ,
\ee
and with the energy $2 p\,  c+ \sigma \, r$ of eq. (\ref{ultrarelativistic})
the linear trajectories can be derived.
The linear trajectories for the angular excitations 
can be derived from the circular classical orbits,
\bea
&& \left\{ \begin{array}{rcl} 2 \left( 2 \pi \, {r \over 2} \, p \right) &  \simeq & l \, h \\  
\sigma &=& {\ \scriptstyle c \ \over  { \scriptstyle r \over  \scriptstyle 2} } p  \end{array} \right.
\nonumber \\
&\Rightarrow & l \simeq { 1 \over 8 \sigma  c  \, \hbar} E^2 \ , 
\label{Bohr angular}
\eea
where we also used the centripetal acceleration.
The linear trajectories corresponding to radial 
excitations can be derived 
\cite{Arriola:2006sv}
from the linear classical orbits with $\mathbf L=0$, 
\bea
&& \int_{-E/\sigma}^{E/\sigma}{ E-\sigma |r| \over 2 c} dr \simeq n \, h 
\nonumber \\
&\Rightarrow & n \simeq { 1 \over 4 \pi \sigma c \, \hbar } E^2 \ .
\label{Bohr radial}
\eea
Thus, in our units of $\hbar=c=\sigma=1$ we get for the Regge slopes respectively,
\bea
\alpha &=& { 1 \over 8} \ ,
\nonumber \\
\beta &=& { 1 \over 4 \pi} \ ,
\label{theor slopes}
\eea
in excelent agreement with the slopes of Figs. \ref{angular Regges} and  \ref{radial Regges},
and similar to a recent Bethe-Salpeter calculation
\cite{Wagenbrunn:2007ie}.

Although the simplified model of eq. (\ref{ultrarelativistic}) agrees qualitatively 
with the experiment, both in the chiral degeneracy and in the linear Regge trajectories, 
there are quantitative discrepancies. In particular, experimentally, 
the radial and agular slopes defined in
eq. (\ref{Regge slopes}) should be almost identical
\cite{Bugg:2004xu},
\bea
\alpha_{\mbox{exp}} &=& 0.877 \ \mbox{GeV}^{-2} \ ,
\nonumber \\
\beta_{\mbox{exp}} &=& 0.855 \ \mbox{GeV}^{-2} \ ,
\eea
while our theoretical slopes, depicted in Figs. \ref{angular Regges} and 
\ref{radial Regges} and semi-classically computed in eq. (\ref{theor slopes}), 
differ by $\pi / 2$. 
Moreover the meson masses in Table \ref{ultrarelativistic meson} 
do not exactly comply with the large degeneracy emerging in the observations of Bugg.

%

\section{Searching for a principal quantum number with classical closed orbits}

To better model the large degeneracy, we enlarge our class of time-like central 
potentials, searching for the best possible potential of this class. 
Actually the model of Section III, with a linear potential only, is 
oversimplified. To fit the correct positive intercepts $\alpha_0$ and $\beta_0$ 
of eq. (\ref{Regge slopes}), it is standard in quark models to include in the 
potential a constant negative energy shift and a negative short range Coulomb 
potential. 

To exactly reproduce the large degeneracy, we ask for a spectrum with a principal 
quantum number, similar to the non-relativistic
spectra of the Coulomb potential or of the harmonic oscillator potential, where
the principal quantum numbers are respectively $n+l+1$ and $2 n + l + {3 \over 2}$.
The difference here is that our kinetic energy is the ultrarelativistic one $T=2 \, p \, c$ 
whereas in the non-relativistic case $T={p^2 \over 2 \mu}$.

Notice that in the non-relativistic case, classical closed orbits coincide with a
quantum principal number. We can also address this problem 
searching for classical closed orbits with an ultrarelativistic kinetic energy.
When all the classical orbits are closed then the Hermann-Jacobi-Laplace-Runge-Lenz 
vector is conserved, because this vector does not precess. In the Hamilton-Jacobi formalism, 
this vector commutes with the hamiltonian. The same commutation then also occurs in the quantum 
Schr\"odinger formalism, the formalism we are using now. Then a larger symmetry group, including 
the angular momentum and the Hermann-Jacobi-Laplace-Runge-Lenz vector exists. Finally this 
implies that a principal quantum number exists. 

Thus, rather than solving the ultrarelativistic Schr\"odinger equation for all the
infinite possible central potentials, we prefer to extend the Bertand theorem techniques
to the search of classical closed orbits with the ultrarelativistic kinetic energy.
For simplicity we consider a kinetic energy $T= p \, c$ and a general potential
$V(r)$, used for a single particle in a central potential, comparable to our two-body
problem in the centre of mass frame.

%

Let us consider classical planar trajectories for an ultrarelativistic quark
with the speed of light, and with momentum, 
\bea
\mathbf p &=& {p \over c} \mathbf v
\nonumber \\
&=& {p \over c} \, \dot r \, \widehat e_r + {p \over c} r \, \dot \theta \, \widehat e_\theta \ ,
\eea
subject to a central force,
\be
\mathbf F = - {d \over d r} V  \, \widehat e_r \ ,
\ee
where the notation is obvious. In the plane we have two constants, the angular
momentum $L$ and the energy $E$,
\bea
\mathbf L &=& {p \over c} \, r^2 \, \dot \theta \, \widehat e_\perp \ \ ,
\nonumber \\
E &=& p  \, c + V(r) \  \ .
\eea
Then Newton's law produces the equation for the radius as a function of time,
\be
\left( {E-V \over c^2} \dot r \right)^{\displaystyle \dot{}} - { L^2 \over {E-V \over c^2} r^3} = - {d \over d r} V \ \ .
\ee
To study the condition to get closed orbits it is convenient to relace the variable
time $t$ by the polar coordinate $\theta$ and the function $r$ by it's
inverse $u=1/r$. Then the equation simplifies to,
\bea
{d^2 \over d \theta^2} u +u &=& J(u)
\nonumber \\
J(u) &=&- {E-V(1/u) \over c^2 L^2}  {d \over d u} V(1/u) \ \ .
\label{J(u) eq}
\eea
To get the classical case one only has to replace in the
right hand side of eq. (\ref{J(u) eq}) the factor ${ E-V(1/u) \over c^2} \rightarrow m $.
Thus the ultrarelativistic case has two independent constants
$E$ and $cL$ while the classical case has only one constant
$m/L$.

The theorem of Bertrand can be addressed considering a
trajectory $u$ close to the circular trajectory $u_0$,
\be
u = u_0+ \eta \ \ ,
\ee
where we fix the angular momentum $L$ and the trajectory is defined by the
inverse radius $u$ and derivative at $\theta=0$ and by the energy $E$.
Defining,
\be
\beta^2= 1- {d \over d u}J|_{u=u_0}  \ \ ,
\ee
we get, to leading order in $\eta$,
\bea
{d^2 \eta \over d \theta^2}+ \beta^2 \eta &=& 0  
\nonumber \\
\eta &=& h_1 \,  \cos \beta (\theta-\theta_0) \ \ .
\eea
Thus for very small perturbations to the circular orbit, the condition
for a closed orbit is that $\beta$ is an integer number (if we
want the orbit to close right after one turn) or rational (if we want
it to close after a finite number of turns).
Importantly this implies that $\beta$ is constant, since it can't
change continuously from one orbit with $u_0$ to the next one.
This restricts the class of possible potentials,
\bea
 {d \over d u}J|_{u=u_0} &=& 1 - \beta^2
\nonumber \\
&=& { f^2({1 \over u_0}) \over c^2 L^2 {u_0}^2} - 2 + { u_0 \over f_0} {d f ({1 \over u_0})\over d u_0}
\label{close to circular}
\eea
and the problem here is that, unlike in the non-relativistic limit,
the equation still depends on the parameter $cL$, and thus the closed
orbits are possible, but there is no potential for which all orbits are closed,
since the closing depends on $cL$. 

In the non-relativistic case it is well known that this problem has two solutions
because the first term in the right-hand-side is absent. Then the solutions of 
eq. (\ref{close to circular}), producing closed orbits close to the circular one, 
are the power law potential with 
$V(r) \alpha r^{\beta^2-2}$. For instance the natural $\beta$ correspond to the powers 
-1, 2, 7, 14 ... For more general orbits, we can go up to the third order in the Fourier 
series for $\eta$,
\be
\eta = h_0+h_1 \,  \cos \beta \theta +h_2 \,  \cos 2 \beta \theta  +h_3 \,  \cos 3 \beta \theta
\ee
and this already produces Bertrand's theorem, stating that the closed orbit condition is, 
\be
\beta^2 ( \beta^2 -1) ( \beta^2-4)=0
\ee
This includes only the Coulomb and the harmonic oscillator cases, which indeed have all 
orbits simply closed for a non-relativistic kinetic energy. 

Again, for an ultrarelativistic kinetic energy there is no potential with all orbits closed.

\section{Conclusion}

Here we study possible quark models for the large degeneragy in the
meson spectra reported by Bugg \cite{Bugg:2004xu}.

We start with a semi-relativistic chiral invariant quark model, with relativistic 
kinetic energy, with negative energy components, but with an instantaneous potential. 
For excited states all the different spin-tensor potentials 
\cite{Bicudo_scalar,Llanes-Estrada_hyperfine,Villate}
merge and we arrive at a schr\"odinger-like ultrarelativistic linear potential quark model 
with a simple $\mathbf J ^2$ dependence. Although the chiral degeneracy is expected to only 
dominate the high excitations, this schr\"odinger-like quark model is conveniently
simple to address the large degeneracy.

The most ambitious approach to the large degeneracy consists in asking for a model with a 
principal quantum number. Notice that when the corresponding classical orbits are all closed, 
the Hermann-Jacobi-Laplace-Runge-Lenz vector is conserved. 
In quantum mechanics this corresponds to a larger symmetry, both angular and 
dynamical, and to a single principal quantum number that encompasses both the angular and 
radial excitations. Unfortunately we find that this is not possible in our ultrarelativistic 
and instantaneous framework.

Another possible approach is the one of an approximate large degeneracy. 
The linear and radial excitations are so regular in the framework of the
ultrarelativistic and equal time quark model that approximate patterns
occur in the spectrum. Notice that in eq. (\ref{theor slopes}),
in Figs. \ref{angular Regges} and \ref{radial Regges}, and in 
Table \ref{ultrarelativistic meson}, an increment of 3 in $j$ is 
approximately equivalent to an increment of 2 in $n$. 
For instance the meson with 
$j=5, \, n=0$ has a similar mass to the one with $j=2, \, n=3$. Notice that the 
$j=5$ state may have $l=4,\, 5,$ or 6, while the $j=2$ state may have $l=1,\, 1,$ 
or 3. we get an approximate but relatively large degeneracy of states with 
different $l$ ranging from 1 to 6. To agree with the experiment, we only need to 
have an increment in $n$ much closer to the increment in $j$. 

Thus, both for the principal quantum number, and for an approximate
degeneracy, a departure from the ultrarelativistic and equal time 
quark model is necessary. It is plausible that either retardation, string effects, 
or coupled channel effects should improve the theoretical model.
For instance Morgunov {\em et al.}
\cite{Morgunov:1999fj}
already showed that including the angular momentum of the rotating linear confining string changes the slopes of the angular Regge trajectories.
Moreover, if one could succeed to include retardation properly in the confining 
quark model, then one might search again for the possible existence of a principal quantum number. And coupled channels might also affect differently 
the angular and radial excitations of the spectrum.

In any case the large degeneracy apparent in the data analysis of Bugg,
supported by the solid pattern of linear angular and radial Regge trajectories, 
remains quite plausible from a theoretical perspective. More experimental data 
on the remarkable patterns of excited resonances is welcome.

\acknowledgements

PB tanks David Bugg and Leonid Glozman for stressing the 
remarkable experimental excited meson data, Pedro Sacramento 
and Jorge Santos for mentioning Bertrand's theorem, and Manuel 
Malheiro for remarking some years ago the approximate
symmetry in the spectrum of chiral invariant quark models. 
PB was supported by the FCT grants POCI/FP/63437/2005 and 
POCI/FP/63405/2005.


\end{document}